\documentclass[letter,12pt]{article}
\usepackage{amsmath,amssymb,amsthm,mathtools}
\usepackage{natbib}
\usepackage{graphicx}
\usepackage[nodisplayskipstretch]{setspace} %
\usepackage{enumitem}
\usepackage{fullpage,rotating,afterpage}
\usepackage{microtype} %

\usepackage[T1]{fontenc} 
\usepackage[utf8]{inputenc} 

\usepackage{wrapfig,boxedminipage}

\usepackage{tikz}
\usepackage{pdflscape}
\usepackage{booktabs}
\usepackage[justification=justified]{caption}
\usepackage[flushleft]{threeparttable}
\usepackage{multirow}

\newcommand{\E}{\mathbb{E}}

\numberwithin{equation}{section}

\usepackage{verbatim,color}

\def\boxit#1{\vbox{\hrule\hbox{\vrule\kern6pt
          \vbox{\kern6pt#1\kern6pt}\kern6pt\vrule}\hrule}}

\newcommand{\captionfonts}{\small}

\makeatletter  % Allow the use of @ in command names
\long\def\@makecaption#1#2{%
  \vskip\abovecaptionskip
  \sbox\@tempboxa{{\captionfonts #1: #2}}%
  \ifdim \wd\@tempboxa >\hsize
    {\captionfonts #1: #2\par}
  \else
    \hbox to\hsize{\hfil\box\@tempboxa\hfil}%
  \fi
  \vskip\belowcaptionskip}
\makeatother   % Cancel the effect of \makeatletter
 \usepackage{titlesec}
 \titleformat{\section}[block]{\centering\normalfont}{\thesection.}{0.5em}{\uppercase }
 \titleformat{\subsection}[runin]{\normalfont}{\thesubsection.}{0.4em plus .1em minus .2em}{\bfseries}[.]
 \titleformat{\subsubsection}[runin]{\normalfont}{\thesubsubsection.}{0.4em plus .1em minus .2em}{\it}[.]
 \titlespacing*\section{0pt}{18pt plus 4pt minus 2pt}{5pt plus 2pt minus 2pt}
 \titlespacing*\subsection{0pt}{10pt plus 2pt minus 1pt}{5pt plus 2pt minus 2pt }
 \titlespacing*\subsubsection{0pt}{4pt plus 1pt minus 1pt}{5pt plus 2pt minus 2pt}
\usepackage{chngcntr}
\usepackage{apptools}
\AtAppendix{\counterwithin{lemma}{section}}
 
 \makeatletter
 \def\mythanks#1{%
 	\protected@xdef \@thanks {\@thanks \protect \footnotetext [\the \c@footnote ]{#1}}%
 }
 \makeatother

\title{\bfseries Combining Population and Study Data for Inference on Event Rates\mythanks{ This Version: \today.}}

\author{ Christoph Rothe\footnote{Department of Economics, University of Mannheim, 68131 Mannheim, Germany.
E-Mail: rothe@vwl.uni-mannheim.de. Website: http://www.christophrothe.net. 
 }}
\date{}

\begin{document}

\pagestyle{plain}

\newtheorem{theorem}{Theorem}
\newtheorem{definition}{Definition}
\newtheorem{lemma}{Lemma}
\newtheorem{assumption}{Assumption}
\theoremstyle{definition}
\newtheorem{example}{Example}
\newtheorem{remark}{Remark}

\newtheorem{assumptionLL}{Assumption}
\renewcommand\theassumptionLL{LL\arabic{assumptionLL}}

\bibliographystyle{ecta}
 
%===Deckblatt=======================================================%
\maketitle 
%===================================================================%

\onehalfspacing

\section{Introduction}

In a recent study, \citet{Streeck2020}  estimate the infection fatality rate  (IFR) of  SARS-CoV-2 infection in a  German town that experienced a super-spreading event in mid-February 2020. The study features prominently in Germany's  current political discussion, and has been covered extensively by major German and international news outlets.
Several newspaper articles raised the question, however, whether the study reports an accurate 
confidence interval (CI) for its IFR estimate.

To explain the issue, consider a stylized version of the setup in \citet{Streeck2020}.
There is a population of total size $N_T$, in which $N_I$ individuals are infected,
and $N_D$ units have died from the infection. The values $N_T$ and $N_D$ are known from administrative records,  but $N_I$ is not directly observed. Instead, the researcher collects a  random sample
of $N_S$ individuals, and observes that $N_P$ of them test positive for the disease. If 
the test  is always accurate, the  IFR can then be estimated by
$$ \widehat{\theta} = \frac{N_D}{\widehat N_I}, \quad\textnormal{ where }\quad  \widehat N_I = \frac{N_P}{N_S}\cdot N_T$$
is an estimate of the number of infected units in the population.
Now, the CI for the IFR reported in \citet{Streeck2020} only takes the sampling uncertainty
about $\widehat N_I$ into account, but treats the number of deaths $N_D$ as fixed.
The question is whether doing so is appropriate, or if $N_D$ should be treated as random.
We argue that the answer depends on whether $\widehat{\theta}$ is interpreted as an estimate of the IFR
among the $N_I$ infected individuals, or an estimate of the IFR among  all $N_T$ members of the population.

To clarify this point, we postulate the existence
of vectors  $\mathbf{D} =(D_1,\ldots, D_{N_T})$, $\mathbf{I} =(I_1,\ldots, I_{N_T})$ and $\mathbf{S} =(S_1,\ldots, S_{N_T})$, with
 $D_j\in\{0,1\}$  an indicator for the (possibly counterfactual) event that the
$j$th individual in the population would have died in the study period if s/he had been infected
with SARS-CoV-2, $I_j\in\{0,1\}$  an indicator for the
$j$th individual  actually being infected, and $S_j\in\{0,1\}$  an indicator for the
$j$th individual being included in the sample. 
These indicators are in principle unobserved, and such
that $$N_S = \sum_{j=1}^{N_T}S_j, \quad N_P = \sum_{j=1}^{N_T}S_jI_j, \quad N_I = \sum_{j=1}^{N_T}I_j, \quad N_D = \sum_{j=1}^{N_T}I_jD_i, \quad N_{D,C} = \sum_{j=1}^{N_T}D_j,$$
with the last term being a new notation for the counterfactual number of deaths one would have observed if the entire population had been infected at the time of the study.

We consider $\mathbf{D}$ to be a fixed feature of the population, and both $\mathbf{S}$ 
and $\mathbf{I}$ to be random vectors whose distribution is determined by the sampling design used in the study and the process that governs the spread of the infection, respectively. This means that $N_I$ and $N_D$ are also random, through
their dependence on $\mathbf{I}$. 
 There are then two plausible candidates for the parameter of interest : the IFR among the individuals that were infected 
 at the time  of the study, given by 
 $$ \theta_1 = \frac{N_D}{N_I},$$
 and the IFR for the entire population, given by 
 $$\theta_2  = \frac{N_{D,C}}{N_T}.$$
 
Now consider a CI that only accounts for the
uncertainty in $\widehat{\theta}$ through its dependence on $\widehat{N}_I$,
which can be
obtained by scaling a conventional $(1-\alpha)$ CI for the proportion of infected individuals. 
For example,
if $(L_\alpha,U_\alpha)$ is a conventional $(1-\alpha)$  Clopper-Pearson CI for the proportion $N_I/N_T$, such a $(1-\alpha)$ CI
 is given by
$$\mathcal{C}^\alpha_1 = \left(\frac{N_D}{N_T\cdot L_\alpha}, \frac{N_D}{N_T\cdot U_\alpha}\right).$$
This type of CI is reported in \citet{Streeck2020}, and it is easily seen to have correct coverage for $\theta_1$ conditional on $\mathbf{I}$, and therefore it  must also have correct coverage
unconditionally: 
$$P( \theta_1 \in \mathcal{C}^\alpha_1|\mathbf{I}) = 1-\alpha \Rightarrow 
P( \theta_1 \in \mathcal{C}^\alpha_1) = 1-\alpha.$$ 
In that sense, the CI in \citet{Streeck2020} is not wrong, but it is
a CI for a very particular target parameter.

In general, inference on $\theta_2$ is going to be more practically relevant since IFR estimates are typically used to design policy measures that affect the entire population. 
The CI $\mathcal{C}^\alpha_1$ clearly does not have correct coverage for  $\theta_2$ 
though, with or without conditioning on $\mathbf{I}$. 
Intuitively, an appropriate CI for $\theta_2$ should be wider than $\mathcal{C}^\alpha_1$, but it is not immediately obvious how such a CI should be constructed. In the remainder of this note, we propose two approaches that both result in good coverage
properties. 
To avoid modeling the number of infections, we seek CIs $\mathcal{C}_2^\alpha$ that are valid conditional on $N_I$,
$$P( \theta_2 \in \mathcal{C}^\alpha_2|N_I) \approx 1-\alpha,$$ 
and any CI that has such approximately correct conditional coverage must again
also have approximately
correct  unconditional coverage. Note that the distinction between $\theta_1$ and $\theta_2$ is similar in spirit
to that
of sampling-based and design-based uncertainty in \citet{abadie2020sampling}, but
the  details
of their framework are very different from ours.

\begin{comment}
Another possibility is to treat the numbers of deaths and infected individuals as random quantities whose realization at the time and location of the study are uncertain before the start of the pandemic,
and define the target of $\widehat{\theta}$ as the expected value of their ratio. With this
view, both $N_D$ and $N_I$  random variables, and the target of $\widehat{\theta}$ 
is
$$ \theta_2 = \E\left(\frac{N_D}{N_I}\right).$$
In such a framework, the estimator $\widehat{\theta}$ is subject to both sampling uncertainty
(through $\widehat{N}_I$ in its denominator) and what one might call ``disease progression''  uncertainty (mainly  through $N_D$ in its numerator, but also indirectly through the denominator since the distribution of $\widehat{N}_I$ depends on $N_I$). This view is arguably more appropriate if the goal is to use a
 CI to describe the uncertainty about the IFR in  populations other than the
  one under study, as in such other populations the number of deaths might be different even if
 the number of infections and all other major population characteristics were exactly the same. 
\end{comment}

\section{Assumptions}

We impose the following assumptions for our analysis.

\begin{assumption}
The sampling and infection indicators are independent conditional on $N_I$:
$$\mathbf{S}\bot \mathbf{I}|N_I$$
\end{assumption}

\begin{assumption}
The infection status of each individual is as good as randomly assigned conditional on $N_I$, in the sense that for all $N_T$-vectors $\mathbf{i}=(i_1,\ldots, i_{N_T})$ of dummy variables 	with $\sum_{j=1}^{N_T} i_j = N_I$ we have that:
$$P(\mathbf{I}=\mathbf{i}|N_I) =  {N_T \choose N_I}^{-1}.$$ 
\end{assumption}

\begin{assumption}
The individuals included in the study sample are determined by simple
random sampling independently of $N_I$, in the sense that for all $N_T$-vectors $\mathbf{s}=(s_1,\ldots, s_{N_T})$ of dummy variables with  $\sum_{j=1}^{N_T} s_j=N_S$ we have that
$$P(\mathbf{S}=\mathbf{s}|N_I) =  {N_T \choose N_S}^{-1}.$$
\end{assumption}

Assumption 1 is natural, and likely to hold even unconditionally. It would be violated,
for example, if individuals with knowledge of their infection status
are more or less like to participate in the study. Assumption 2 implies that the individuals infected
at the time of the study are representative for the entire population. This rules out, for example, different age groups being affected more or less
severely over the course of the pandemic.  Note that the ``success'' probability $N_I/N_T$ can be changed to accommodate infection testing
with less than 100\% sensitivity and specificity.
Assumption~3   can easily be adapted if the sample
of $N_S$ individuals is obtained though a different sampling scheme, such as cluster sampling. 
Note that an equivalent definition of  $\theta_2$ under the above assumptions  is given by
$$\theta_2 = \E\left(\frac{N_D}{N_I}\right),$$
so that this parameter can be interpreted as the ``average'' IFR, where the averaging
is done with respect to the distribution of  $\mathbf{I}$. This representation
also makes it more apparent that $\widehat{\theta}$ is actually a suitable estimate of $\theta_2$.

Since  $\widehat{\theta}$ depends on $\mathbf{S}$ and $\mathbf{I}$ through $N_P$ and $N_D$ only, it is also useful to state the implications of the above assumptions for the joint
distribution of the latter two quantities conditional on $N_I$. Simple calculations show that this
joint conditional distribution corresponds
to two independent binomials:
$$N_P \bot N_D|N_I, \qquad N_P|N_I\sim \textnormal{Binomial}\left(N_S, \frac{N_I}{N_T}\right), \qquad 
N_D|N_I  \sim \textnormal{Binomial}\left(N_I , \theta_2 \right).$$
These distributions should be kept in mind for the following arguments.

\section{Confidence Sets}

Consider a  test of the null hypothesis $H_0: \theta_2 = \theta^o$ that uses the estimated IFR $\widehat{\theta}$ as the test statistic. We propose to construct $(1-\alpha)$ CIs for $\theta_2$ by collecting all values of $\theta^o$ for which the  $p$-value of such a test  is less than $\alpha$. With conditioning on   $N_I$, the number of infections  effectively becomes a nuisance
parameter in this testing problem; and since $N_I$ is unknown   no exact $p$-value is feasible in this setup. However,  we can still use existing statistical approaches to obtain CIs with good coverage properties. We specifically consider one based on the parametric bootstrap,
and one based on varying $N_I$ over a ``large'' preliminary CI.

%$$\widehat{\theta} - \theta_2 = \frac{N_D}{\widehat{N}_I} - \frac{N_D}{N_I} + \frac{N_D}{N_I}  -\frac{\E\left(\left.N_D\right|N_I\right)}{N_I} \approx -\frac{N_D}{N_I^2}(\widehat{N}_I-N_I) + \frac{N_D}{N_I}  -\frac{\E\left(\left.N_D\right|N_I\right)}{N_I} $$
%$$\V(\widehat{\theta} - \theta_2|N_I) \approx ...$$

To describe these two approaches in our context, we introduce some notation. 
For constants $n_I$ and $\theta^o$, let  $N_P^*$ and $N_D^*$ be independent random variables
that each follow particular binomial distributions that only depend on the constants and other
observable quantities: 
$$N_P^* \bot N_D^*, \quad N_P^* \sim \textnormal{Binomial}\left(N_S, \frac{n_I}{N_T}\right),
\quad N_D^* \sim \textnormal{Binomial}\left(n_I, \theta^o \right). $$
We also put $\widehat{N}^*_I =N_T N_P^*/N_S$, and denote the CDF
of the ratio $N_P^*/\widehat{N}^*_I$ by
$$G(c|n_I,\theta^o) = P\left( \frac{N_D^*}{\widehat{N}_I^*} \leq c\right).$$
There is no simple closed form expression for this distribution function, but it can  easily be computed through standard numerical methods for any value of the constants  $n_I$ and $\theta^o$.
For example, one can compute $G(c|n_I,\theta^o)$ to desired accuracy by simulating a sufficiently large
number of draws from the distribution of $(N_P^*,N_D^*)$, and then taking the empirical CDF
of the resulting realizations of   $N_P^*/\widehat{N}^*_I$.

The function $G(c|N_I,\theta_2)$ is the CDF of $\widehat{\theta}$ conditional on $N_I$
under the statistical model described above, and  $G(c|N_I,\theta^o)$ is the CDF under
 $H_0: \theta_2 = \theta^o$. If $N_I$ was observed, an equal-tailed $p$-value for a test of $H_0$
 based on $\widehat{\theta}$ would be given
 by $$p(\theta^o,N_I) = 2\min\left\{\widehat G(\widehat{\theta}|N_I,\theta^o), 1-\widehat G(\widehat{\theta}|N_I,\theta^o)\right\}.$$
 
Using a ``plug-in'' or  parametric bootstrap approach \citep[e.g.][]{horowitz2001bootstrap,hall2013bootstrap}, we can substitute the estimator $\widehat{N}_I$
into the $p$-value  formula 
 to construct a feasible  CI for $\theta_2$:
$$\mathcal{C}_{2,PB}^{\alpha} = \{\theta^o: p(\theta^o,\widehat{N}_I) \geq \alpha \}.$$
This CI is easily seen to have correct asymptotic coverage of $\theta_2$ conditional on $N_I$
under any sequence for which $\widehat N_I / N_I = 1 + o_P(1)$. That is, it holds that
$$P( \theta_2 \in \mathcal{C}^\alpha_{2,PB}|N_I) = 1-\alpha + o_P(1) \quad\textnormal{ if }\quad\widehat N_I / N_I = 1 + o_P(1).$$ 
 If the sample size $N_S$  is rather large, it can be reasonable to treat $\widehat N_I$ as a consistent
estimate of $N_I$, in which case the above result implies that $\mathcal{C}_{2,PB}^{\alpha}$ has approximately correct finite sample coverage of $\theta_2$.

If the goal is to have a CI with guaranteed finite sample coverage, a different method can be used to
compute a $p$-value. Let $[ L_\beta; U_\beta ]$ be a standard $(1-\beta)$ Clopper-Pearson CI for the share $N_I/N_T$ of infected individuals in the population, so that $\mathcal{C}^\beta = [N_T L_\beta; N_T U_\beta ]$ is a $(1-\beta)$ CI for the number of infections $N_I$, for some $\beta$    substantially smaller than $\alpha$.
We can then obtain a new $p$-value by maximizing $p(\theta^o,n_I)$ over $n_I\in\mathcal{C}^\beta$, and correcting the result for the fact that $\beta$ is not zero \citep{berger1994,Silvapulle1996}. This yields the following   CI for $\theta_2$:
$$\mathcal{C}_{2,CS}^{\alpha} = \left\{\theta^o: \sup_{n_I\in \mathcal{C}^\beta} p(\theta^o, n_I) + \beta \geq \alpha \right\}.$$
This CI has conditional coverage of at least $1-\alpha$ in finite samples: 
$$P( \theta_2 \in \mathcal{C}_{2,CS}^\alpha|N_I) \geq 1-\alpha.$$ 
The CI is   conservative, however, in  that the last inequality is generally
strict. Exact coverage only occurs in the unlikely scenario that the supremum
in the definition of the $p$-value is attained  at $N_I$, which happens only if $N_I$ coincides with one of the boundaries of $\mathcal{C}^\beta$.

\section{Numerical Illustration}

We illustrate methods described above with numerical values taken from \citet{Streeck2020}. The town investigated in that study has $N_T = 12,597$  inhabitants, of which $N_D = 7$
died in the study period with a SARS-CoV-2  infection. Out of a sample of $N_S=919$ individuals, $N_P = 138$ tested positive for SARS-CoV-2. This corresponds to an infection rate
of $N_P/N_S = 15.0\%$ in the sample, an estimated $\widehat{N}_I = 1892$ infected
individuals in the population, and an estimated IFR of $\widehat{\theta}=0.37\%$. Setting
$\alpha =.05$ and $\beta=.01$, we obtain the CIs
$$\mathcal{C}_1^{\alpha} = [0.32\%; 0.43\%], \qquad \mathcal{C}_{2,PB}^{\alpha} = [0.16\%; 0.74\%], \qquad \mathcal{C}_{2,CS}^{\alpha} = [ 0.14\%; 0.81\%]. $$
Recall that the first of these CIs has $\theta_1$ as the target parameter, while the latter
two aim for coverage of $\theta_2$. As expected, the   latter two CIs are substantially wider than the first. We would argue that they are also more appropriate measures of uncertainty about the IFR estimate, since this
quantity is used to design policy measures that affect the entire population.

We note that \citet{Streeck2020} report an estimated 1,956 infected individuals, an IFR of .36\%, and a  CI for the IFR of $[0.29\%; 0.45\%]$. These results differ from the $\widehat{N}_I$, $\widehat{\theta}$ and $\mathcal{C}_1^{\alpha}$ given above
for two reasons: first, \citet{Streeck2020} apply an adjustment factor to the raw infection
rate in their sample to account for the sensitivity and specificity of their test for SARS-CoV-2 infection;
and second, their sample is generated through a form of cluster sampling, which leads to a slightly wider CI relative to simple random sampling. Such adjustments should also slightly widen our CIs for $\theta_2$.

\section{Discussion}

While this note is motivated by research on the current SARS-CoV-2 pandemic, the CIs proposed here could also be used in other contexts in which researchers want to combine sample and population
data in a similar fashion. To give an economic example, suppose that there is a group of
individuals that qualify for benefits from  some public program, and that the researcher is interested
in the share of these individuals that actually receive benefits (this share could be small if the
program is not well-known, difficult to apply for, or comes with social stigma).
This then fits into the framework of this note if the number of benefit recipients is known to administrators,
but the number of qualifying individuals needs to be estimated from survey data.

\singlespacing

\bibliography{bibl}    

\end{document}